\documentclass[smallextended]{svjour3}       
\smartqed  
\usepackage{graphicx}
\usepackage{url}
\usepackage[colorlinks,citecolor=blue,linkcolor=blue,urlcolor=blue]{hyperref}
\newcommand{\bl}{\begin{aligned}}
\newcommand{\el}{\end{aligned}}   
\newcommand{\be}{\begin{equation}}
\newcommand{\ee}{\end{equation}}   
\newcommand{\bea}{\begin{eqnarray}}
\newcommand{\eea}{\end{eqnarray}}
\newcommand{\ba}{\begin{array}}
\newcommand{\ea}{\end{array}}

\newcommand{\la}{\left <}
\newcommand{\ra}{\right >}

\begin{document}

\title{Role of atomic vacancies and second-neighbor antiferromagnetic-exchange coupling in a ferromagnetic nanoparticle}

\author{Harun Al Rashid$^1$     \and
        Muskan Sharma$^2$       \and
        Shruti$^2$              \and
        Dheeraj Kumar Singh$^1$}

\institute{
\email{dheeraj.kumar@thapar.edu} \\
$^1$Department of Physics and Material Science, Thapar Institute of Engineering and Technology, Patiala-147004, Punjab, India\\
           $^2$Department of Physics,
Panjab University, Chandigarh }
\date{Received: date / Accepted: date}
\maketitle

\begin{abstract}
Several factors may be responsible for disorder
and frustration in a magnetic nanoparticle, including atomic vacancies on the surface and inside, impurity atoms, long-range magnetic exchange coupling, etc. We use Monte-Carlo simulations within the Heisenberg model to examine the role of randomly distributed atomic vacancies and long-range magnetic-exchange coupling on the temperature-dependent magnetic properties of ferromagnetic nanoparticles. In particular, we study the role of the second-neighbor antiferromagnetic exchange coupling and missing atoms inside the particle resulting in broken nearby bonds. We find that both factors may enhance the superparamagnetic behaviors of such particles.

\end{abstract}

\section{Introduction}
Magnetic nanoparticles have attracted considerable attention in recent times due to their vast applications in various fields ranging from high-density data storage to diverse biomedical uses~\cite{Duguet,Maria,Vatta,Pankhurst,Beydoun}. These particles may exhibit magnetic behavior, which may be remarkably different from their bulk counterparts~\cite{Anwar,Goveas,Kolhatkar,Garcia,Goya}. This is largely due to the finite-size effect characterized by discrete energy levels while the translational symmetry is absent. Additionally, the surface-to-volume ratio continues to increase with a reduction in the size of nanoparticles, which can also have a significant impact on their magnetic properties~\cite{Pozzi,Abbasi}.

The Curie temperature of ferromagnetic (FM) nanoparticles decreases with their size whereas it may either decrease or increase depending on the types of atoms being doped~\cite{CAO,Afify,Bahadur,Layek,Pal,Cong}. Similarly, the coercive field may also decrease or increase with a decrease in size. On the other hand, the magnetization of antiferromagnetic (AFM) nanoparticles below the N\'{e}el temperature is larger than the bulk value because of uncompensated spins on the surface~\cite{Wesselinowa,Montes,Tobia,Skomski,Markovic}.

The surface atoms reside in a highly anisotropic environment because of the absence of several bonds depending on their location on the surface and nature of the surface~\cite{Garanin,Labaye}. The absence of bonds, which is also dependent on the location on the surface and nature of the surface, introduces frustration and hence disorder into the system~\cite{Kodama,Sousa}. Other factors which are also an important role in affecting their magnetic properties are their shape, surface/bulk anisotropy, surface roughness, doping which involves replacing some of the atoms by other magnetic or nonmagnetic atoms etc~\cite{Pastukh,Egbu,Heo,Gomonay}. Besides the surface, the process of synthesization can also create atomic vacancies well inside the nanoparticle, an issue which remains largely unexplored~\cite{Zou,He}. Secondly, a long-range orbital overlap can also lead to long-range magnetic exchange coupling~\cite{Danilyuk}. For instance, in certain material systems, a second-neighbor ferromagnetic or antiferromagnetic exchange or even higher-order magnetic exchange coupling can also be nonnegligible. For a ferromagnetic particle, a second-neighbor ferromagnetic coupling can further strengthen the ferromagnetic order. On the other hand, antiferromagnetic coupling can introduce frustration for the ferromagnetic order.

Earlier works have focused largely on the role of shape and surface roughness~\cite{salazar,Pierce,Razouk} on the magnetic properties of ferromagnetic nanoparticles. However, the atomic vacancies are found to increase in concentration as the size of nanoparticles becomes smaller~\cite{wanga,goyal}. Moreover, in a correlated electron system, wherein $d$ and $f$ orbitals are active with on-site Coulombic interaction being large, the long-range magnetic correlations become important~\cite{agarwal}. In most of the previous works on magnetic nanoparticles, these two factors have largely been ignored. To fill this gap, we explore the role of a finite number of atomic vacancies inside the nanoparticle, which are randomly distributed as well as how the presence of long-range magnetic coupling, in particular, antiferromagnetic affects the finite-temperature magnetic properties. We use Monte Carlo (MC) to simulate the spin of the Heisenberg model while treating them as classical fields. The accuracy of this approach increases with temperature as it incorporates all types of possible thermal fluctuations unlike the mean-field theories, which ignore thermal fluctuations of spins.

\section{Model and Method}
We consider the Heisenberg Hamiltonian to describe the localized spins of a magnetic nanoparticle on a simple cubic lattice structure~\cite{Razouk}. The Hamiltonian is given by
\begin{equation}
 	H = -\sum_{i,\delta}^{N} J_{i,i+\delta}{\bf S_i}\cdot{\bf S_{i+\delta}} -
  K_b\sum_i^N{ S}^2_{iz} - K_s\sum_{i}^{N}({\bf S}\cdot\hat{n})^2
\end{equation}
The first term describes the first- and second-neighbor exchange interaction between the spins. A positive/negative $J_{i,i+\delta}$ drives the system towards FM/AFM. A positive $J$ originates from the standard exchange interaction or through double-exchange interaction in the ferromagnetic-kondo lattice model~\cite{kondo,pandey,dks1,dks2}. On the other hand, a negative $J$ naturally follows from the second-order perturbation theory~\cite{anderson} or appears naturally in the Monte-Carlo simulations\cite{harun1,harun2} within the Hubbard model, and it is often referred to as ``superexchange interaction''. The second term originates from the uniaxial anisotropy responsible for favoring the spin alignment along the easy axis, which in this work is considered to be the $z$ axis. The third term incorporates the surface anisotropy favoring the spin alignment perpendicular to the surface. ${\bf n}$ is the unit vector perpendicular to the surface. These anisotropies are the consequence of the spin-orbit interactions. In the following we set $J_b$, the nearest-neighbor exchange coupling parameter, which is positive, as the unit of energy. It may be noted that $J_b \sim$ 50meV~\cite{liu} is the largest interaction parameter in the thoery. We fix the bulk anisotropy parameter to be $K_b = 0.1J_b$ whereas the surface anisotropy parameter is set to be $K_s = 0.2J_b$ as surface anisotropy is comparatively larger. For simplicity, the magnitude of the spin vector $\{{\bf S}_i\}$ is fixed to be unity.

Through the MC simulation, an equilibrium configuration corresponding to a minimum energy configuration at a given temperature is obtained. The classical spin variables $\{{\bf S}_i\}$ are annealed with the help of Metropolis algorithm by cooling down the system from the higher temperature to the lower temperature~\cite{Zhang}.
Unlike an itinerant model such as the Hubbard model, where the simulation process may be complex, it is rather simple for a localized model and involves the following steps as follows: first, we choose a site ${\bf i}$ with spin $({\bf S}_i)$ and calculate energy $(E_i)$ of the system. Thereafter, the orientation of the spin of the chosen site is randomly updated to ${\bf S}^\prime_i$, and the energy of the system $E_f$ is calculated. The energy difference $({\Delta} E = E_f - E_i)$ for the spin configurations before and after update is obtained. If ${\Delta}E < 0$, the update is accepted. If ${\Delta}E \geq 0$, the update is accepted only after comparing the Boltzmann thermodynamic probability factor $P(E) \propto$ exp$(-{\Delta}E/k_BT)$ with a random number $0<r<1$. If $P(E) > r$, the update is accepted, otherwise, it is rejected. A cycle, in which all the atomic sites of the nanoparticle go through this process, is called one Monte-Carlo sweep (MCS). In the simulation of a given system without disorder and frustration, $\sim$2000 Monte Carlo sweeps have often been considered sufficient as noted in various studies. This is because, in such system, random moves to sample the configuration space (change in spin orientation in the current work) in the MC simulations leads to a more predictable trajectory towards the equilibrium. In the current work, to be on safer size, a large number $\sim 6000$ of sweeps is considered in order to reach the equilibrium configuration, which is because of the presence of second-neighbor antiferromagnetic exchange coupling generating frustration and atomic vacancy sites responsible for the disorder~\cite{patel}. Both the factors have ability to slow down the process of equilibration.

The observables that we use to study thermodynamical magnetic properties are magnetization given by
\begin{equation}
 \langle M_z \rangle  = \la  \frac{1}{N} \sum_i{m_{iz}} \ra
\end{equation}
 and magnetic susceptibility given by
\begin{equation}
	\chi(T) = N \times \left(\la{M_z^2}\ra - {\la{M_z}\ra}^2\right)/k_BT.
\end{equation}
Here,
\begin{equation}
	\langle M^2_z \rangle  = \la  \frac{1}{N}\sum_i{m_{iz}^2} \ra
\end{equation}
with $m_{ix}$, $m_{iy}$, and $m_{iz}$ being the components of the magnetic moments at a given temperature $T$. $N$ is the total number of sites or spins in the system under consideration. The signature of the onset of FM order and the Curie temperature $(T_c)$ can be estimated from the temperature-dependent behavior of magnetization, the latter starts to rise on crossing the Curie temperature in the cooling process.

In cases, where we focus on the role of vacancies in the system, in addition to the thermal averaging, an averaging over different impurity configurations for a given fraction of atomic vacancies is carried out. In this work, the averaging over atomic vacancies is done over five different configurations to obtain a robust result. Throughout the current work, we assume the nanoparticle to have a cubical shape. Unless otherwise indicated, the particle size is 16 $\times$ 16 $\times$ 16. If the interatomic distance is 0.4nm then the particle will have an edge of length 6.4nm.
\section{Results and Conclusions}

\begin{figure} []
    \centering
     \includegraphics[scale = 1.0, width = 12.2cm]{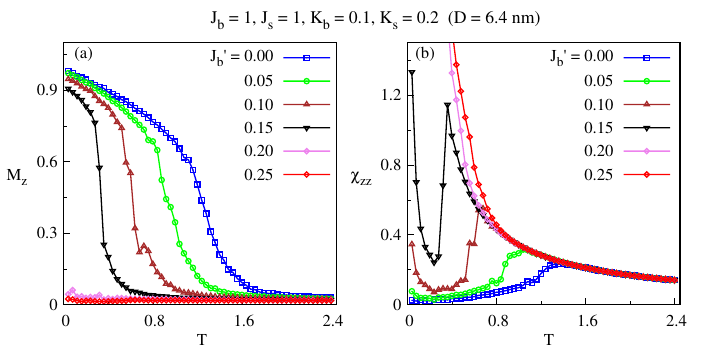}
    \caption{(a) Magnetization and (b) susceptibility as a function of temperature for different second-neighbor antiferromagnetic couplings. }
    \label{1}
\end{figure}
\begin{figure} []
    \centering
     \includegraphics[scale = 1.0, width = 8.0cm]{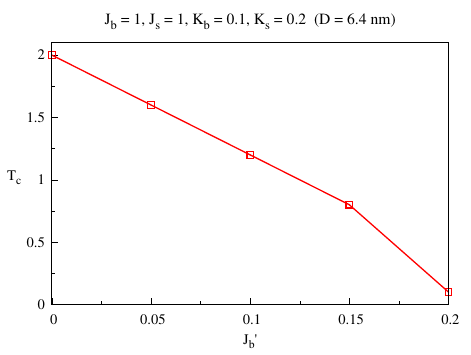}
    \caption{Variation of critical temperature $T_c$ as a function of second-neighbor antiferromagnetic exchange coupling ($J_b'$).}
    \label{2}
\end{figure}
\begin{figure} []
    \centering
     \includegraphics[scale = 1.0, width = 12.2cm]{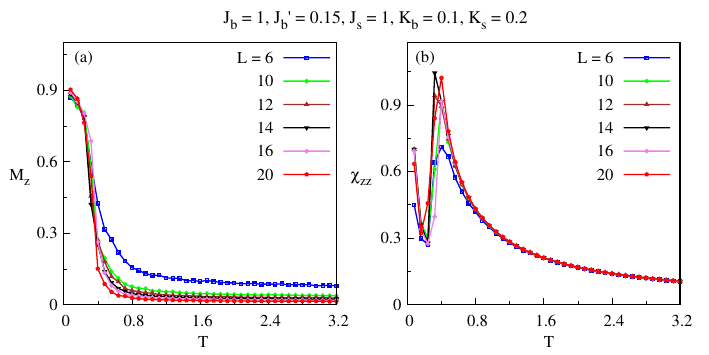}
    \caption{Finite-size effect on the behavior of (a) magnetization and (b) susceptibility as a function of temperature for various edge lengths $L$.}
    \label{3}
\end{figure}
\begin{figure} []
    \centering
     \includegraphics[scale = 1.0, width = 12.2cm]{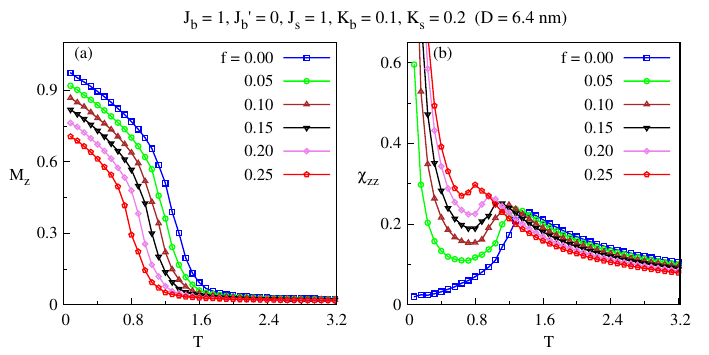}
    \caption{(a) Magnetization and (b) susceptibility as a function of temperature for different fractions of atomic vacancies denoted by $f$.}
    \label{4}
\end{figure}
\begin{figure} []
    \centering
     \includegraphics[scale = 1.0, width = 8cm]{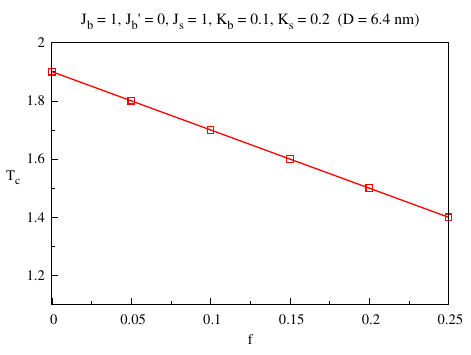}
    \caption{Variation of susceptibility peak as a function concentration of atomic vacancies.}
    \label{5}
\end{figure}
\begin{figure} []
    \centering
     \includegraphics[scale = 1.0, width = 12.2cm]{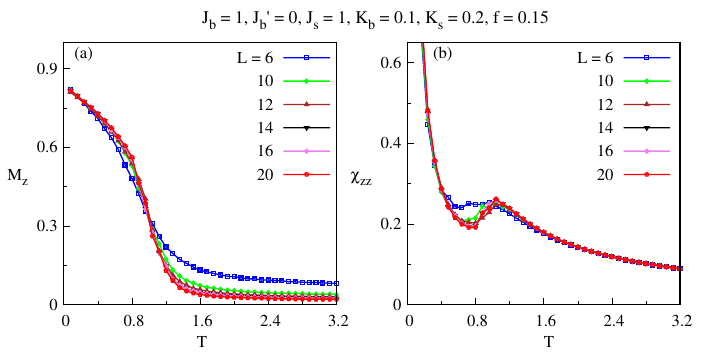}
    \caption{Finite-size effect on the behavior of magnetization and susceptibility as a function of temperature for a given atomic vacancy.}
    \label{6}
\end{figure}
Presence of second-neighbor antiferromagnetic coupling is known to generate frustration in a ferromagnet, {i. e.}, ferromagetic correlation is weakened. Figure~\ref{1}(a) shows the magnetization vs temperature plot for different strength of second-neighbor antiferromagnetic exchange coupling $J^{ \prime}_b$. The saturation magnetization shows only a weak dependence on $J^{ \prime}_b$. It does not show a significant reduction with a rise in $J^{\prime}_b$ until the latter reaches near a critical value $J^{ \prime}_b \sim 0.2$, whereafter it drops at once. Second, the critical temperature, at which, the magnetization starts rising upon lowering of temperature, decreases monotonically until a critical value of $J^{\prime}_b \sim 0.1$. A sharp drop in $T_c$ and only a minor reduction in the saturation magnetization makes the magnetization vs temperature curve a sharp upward turn in the vicinity of $T_c$ as temperature is lowered.

For a smaller value of $J^{\prime}_b$, the magnetic susceptibility shows a behavior as expected, {i.e.}, it is accompanied by a peak near the ferromagnetic-paramagnetic transition $T_c$ (Fig.~\ref{1}(b)). Note that $T_c$ shows almost a linear drop as $J^{\prime}_b$ increases (Fig.~\ref{2}), which indicates that an increasing second-neighbor antiferromagnetic exchange interaction suppresses the onset temperature of long-range ferromagnetic correlation linearly in the nanoparticles~\cite{carvalho}. However, when $J^{\prime}_b$ increases beyond $\sim 0.1$, the susceptibility starts to rise once again after getting peaked at $T_c$ upon decreasing the temperature further. The rise is sharp and without any downturn. In other words, the susceptibility does not vanish at a very low temperature. At the same time, the magnetization does not vanish for $ J^{\prime}_b$ in the range between $\sim$0.1 and $\sim$0.2.
 
For the range of $ J^{\prime}_b$ in between $\sim$0.1 and $\sim$0.2, further rise in magnetic susceptibility at low temperature is a signature of the fact that the bulk anisotropy parameter $K_b = 0.1$ is now not sufficient to hold the magnetic moment of the particle along $z$ direction. In other words, while the nanoparticle is magnetized completely, the magnetic moments cannot align along a particular direction, an indicative of superparamagnetic regime for this range. Although beyond $ J^{\prime}_b$ $\sim$0.2, the saturation magnetization of the nanoparticle itself vanishes. It can be seen that the saturation magnetization is almost independent of particle size whereas differences can be noticed at higher temperature with magnetization being the largest for the smallest particle (Fig~\ref{3}(a)). Nonvanishing of magnetization at higher temperature is a finite-size effect. Moreover, the temperature $T_c$, which corresponds to the onset of ferromagnetic behavior, becomes almost constant (Fig~\ref{3}(b)).

Figure~\ref{4}(a) shows the magnetization as a function of fraction of atomic vacancies distributed randomly inside the particle, where the second-neighbor antiferromagnetic coupling is ignored for simplicity. The effect of concentration of atomic vacancy is explored up to 25\%. Here, both the saturation magnetization as well as the Curie temperature show a continuous drop, when the concentration is increased. A nearly linear drop in the saturation magnetization, when the concentration is increased, results from the fact that the magnetic moments of the missing atoms do not contribute. Second, the broken bonds may also generate magnetic moment fluctuations of thermal origin in contributing to an additional reduction. We find that the Curie temperature decreases a behavior similar to the bulk counterpart~\cite{meyer}. Next, the magnetization vs temperature curve appoaches the saturation magnetization not sharply rather very slowly.

Figure~\ref{4}(b) shows the magnetic susceptibility as a function of temperature. The effect of disorder created by atomic vacancies is  clearly evident in the behavior of the spin susceptibility. The susceptibility does show a peak near the paramagnetic to ferromagnetic transition. However, the peak size shows a continuous decline with temperature and becomes unnoticeable ultimately. The nature of transition changes as quickly as one approaches $f \sim 0.05$. Second, the magnetic susceptibility rises again after getting peaked at $T_c$, a behavior similar to the what is observed when $J^{\prime}_b$ lie in between $\sim 0.1$ and $\sim 0.2$. It can be noticed that even a small atomic vacancy as low as $\sim 5\%$ may bring the nanoparticle to superparamagnetic regime. The shifting of the susceptibility peak towards a smaller value  or decrease in $T_c$ is nearly linear a function of concentration of atomic vacancies (Fig. \ref{5}), which results from the weakening of ferromagneic correlations.
 
We find the behavior of the particles as a function of size similar to the case when $J^{\prime}_b$ was varied (Fig.~\ref{6}). There is no size dependent difference in the saturation magnetization and a difference can be noticed only when temperature is increased. Similarly, one finds that for a given $f$, the small peak in susceptibility becomes further small until it disappears completely.

In conclusion, we have investigated the effect of second-neighbor exchange coupling and atomic vacancies on the temperature-dependent magnetic behavior of ferromagnetic nanoparticles. In order to achieve this objective, we used the Monte-Carlo simulations of the atomic spins within the classical Heisenberg model. Our finding suggests that the frustrations introduced by both factors support the superparamagnetic behavior of the nanoparticle. Understanding the roles of these factors may be beneficial in controlling the superparamagnetic behavior as the latter finds promising applications in targeted drug delivery, magnetic resonance imaging, magnetic hyperthermia, bioseparation, biosensing, etc. On the other hand, it is also known to cause thermal instability or fluctuations in stored information leading to inefficacy in data storage.

\begin{acknowledgements}
 D.K.S. was supported through DST/NSM/R\&D\_HPC\_Applications/2021/14 funded by DST-NSM and start-up research grant SRG/2020/002144 funded by DST-SERB.
\end{acknowledgements}
\section*{Declarations}
The authors declare no conflict of interest. Muskan Sharma and Shruti have made equal contributions.

\section*{Data availability} All the data are available through the manuscript.

\section*{Author contributions} Conceptualization: D K Singh, H A Rashid, Muskan Sharma, Shruti;  Methodology: D K Singh, H A Rashid, Muskan Sharma, Shruti; Computation: H A Rashid, Muskan Sharma, Shruti; Original draft preparation: H A Rashid, Muskan Sharma, Shruti; Finalization of draft (review and editing): D K Singh; Supervision: D K Singh


\begin{thebibliography}{100} 

\bibitem{Duguet} E. Duguet, S. Vasseur, S. Mornet, and J. M. Devoisselle, Nanomedicine {\bf 1}, 157–168 (2006). \url{https://doi.org/10.2217/17435889.1.2.157}

\bibitem{Maria} P. Maria, B. Ingrid, and J. Wolfgang, Chem. Soc. Rev. {\bf 41}, 4306-4334 (2012). \url{https://doi.org/10.1039/C2CS15337H}

\bibitem{Vatta} L. L. Vatta, R. D. Sanderson, and K. R. Koch, Pure Appl. Chem. {\bf 78}, 1793-1801 (2006). \url{https://doi.org/10.1351/pac200678091793}

\bibitem{Pankhurst} Q. A. Pankhurst, J. Connolly, S. K. Jones, and J. Dobson, J. Phys. D: Appl. Phys. {\bf 36} R167 (2003). \url{https://dx.doi.org/10.1088/0022-3727/36/13/201}

\bibitem{Beydoun} D. Beydoun, R. Amal, and G. Low,  J. Nanoparticle Res. {\bf 1}, 439–458 (1999). \url{https://doi.org/10.1023/A:1010044830871}

\bibitem{Welch} C. M. Welch, R. G. Compton, Anal. Bioanal. Chem. {\bf 384}, 601–619 (2006). \url{https://doi.org/10.1007/s00216-005-0230-3}

\bibitem{Anwar} A. Anwar, M. A. Basith, S. Choudhury, Mater. Res. Bull. {\bf 111}, 93-101 (2019). \url{https://doi.org/10.1016/j.materresbull.2018.11.003}

\bibitem{Goveas} L. R. Goveas, K. N. Anuradha, K. S. Bhagyashree, S. V. Bhat, J. Appl. Phys. {\bf 117} 17E111 (2015). \url{https://doi.org/10.1063/1.4913722}

\bibitem{Kolhatkar} A. G. Kolhatkar, A. C. Jamison, D. Litvinov, R. C. Willson, and T. R. Lee, Int. J. Mol. Sc., {\bf 14}, 15977--16009 (2013). \url{ https://doi.org/10.3390/ijms140815977}

\bibitem{Garcia} M. A. Garcia, J. M. Merino, E. F. Pinel, A. Quesada, J. de la Venta, M. L. R. Gonza\'lez, G. R. Castro, P. Crespo, J. Llopis, J. M. Gonz\'alez-Calbet, and A. Hernando, Amer. Chem. Soc. {\bf 7}, 1489-1494 (2007). \url{https://doi.org/10.1021/nl070198m}

\bibitem{Goya} G. F. Goya, T. S. Berquo\', F. C. Fonseca, and M. P. Morales, J. Appl. Phys. {\bf 94}, 3520-3528 (2003). \url{https://doi.org/10.1063/1.1599959}

\bibitem{Pozzi} M. Pozzi, S. J. Dutta, M. Kuntze, J. Bading, J. S. Rüßbült, C. Fabig, M. Langfeldt, F. Schulz, P. Horcajada, W. J. Parak, J. Chem. Educ. {\bf 101}, 3146-3155. \url{https://doi.org/10.1021/acs.jchemed.4c00089}

\bibitem{Abbasi} R. Abbasi, G. Shineh, M. Mobaraki, J. Nanopart Res. {\bf 25}, 43 (2023). \url{https://doi.org/10.1007/s11051-023-05690-w}

\bibitem{CAO} L.-F. Cao, D. Xie, M.-X. Guo, H.S. Park, and T. Fujita, Trans. Nonferrous Met. Soc. China {\bf 17}, 1451-1455 (2007). \url{https://doi.org/10.1016/S1003-6326(07)60293-3}

\bibitem{Afify} M. S. Afify, M. M. El Faham, U. Eldemerdash, W. M.A . El Rouby, and S. I. El-Dek, J. Alloys Compd. {\bf 861}, 158570 (2021). \url{https://doi.org/10.1016/j.jallcom.2020.158570}

\bibitem{Bahadur} N. Bahadur, R. Pasricha, Govind, S. Chand, and R. K. Kotnala, Mater. Chem. Phys {\bf 133}, 471-479 (2012). \url{https://doi.org/10.1016/j.matchemphys.2012.01.068}

\bibitem{Layek} S. Layek and H. C. Verma, J. Magn. Magn. Mater. {\bf 397}, 73-78 (2016). \url{https://doi.org/10.1016/j.jmmm.2015.08.082}

\bibitem{Pal} A. Pal and P. K. Giri, J. Appl. Phys. {\bf 108}, 084322 (2010). \url{https://doi.org/10.1063/1.3500380}

\bibitem{Cong} C. J. Cong, L. Liao, Q. Y. Liu, J. C. Li, and K. L. Zhang, Nanotechnology {\bf 17}, 1520 (2006). \url{https://dx.doi.org/10.1088/0957-4484/17/5/059}

\bibitem{Wesselinowa} J.M. Wesselinowa, J. Magn. Magn. Mater., {\bf 322}, 234-237 (2010). \url{https://doi.org/10.1016/j.jmmm.2009.08.045}

\bibitem{Montes} N. Rinaldi-Montes, P. Gorria, D. Marti\'nez-Blanco, A. B. Fuertes, I. Puente-Orench, L. Olivi, J. A. Blanco, AIP Advances {\bf 6}, 056104 (2016). \url{https://doi.org/10.1063/1.4943062}

\bibitem{Tobia} D. Tobia, E. Winkler, R. D. Zysler, M. Granada, and H. E. Troiani
Phys. Rev. B {\bf 78}, 104412 (2008). \url{https://doi.org/10.1103/PhysRevB.78.104412}

\bibitem{Skomski} R. Skomski, B. Balamurugan, P. Manchanda, M. Chipara and D. J. Sellmyer,  IEEE Trans. Magn. {\bf 53}, 1-7 (2017). \url{https://doi.org/10.1109/TMAG.2016.2601019}

\bibitem{Markovic} V. Markovich, I. Fita, A. Wisniewski, D. Mogilyansky, R. Puzniak, L. Titelman, C. Martin, and G. Gorodetsky, Phys. Rev. B {\bf 81}, 094428 (2010). \url{https://doi.org/10.1103/PhysRevB.81.094428}

\bibitem{Garanin} D. A. Garanin and H. Kachkachi, Phys. Rev. Lett. {\bf 90}, 065504 (2003). \url{https://doi.org/10.1103/PhysRevLett.90.065504}

\bibitem{Labaye} Y. Labaye, O. Crisan, L. Berger, J. M. Greneche, J. M. D. Coey, J. Appl. Phys. {\bf 91}, 8715-8717 (2002). \url{https://doi.org/10.1063/1.1456419}

\bibitem{Kodama} R. H. Kodama, A. E. Berkowitz, E. J. McNiff, Jr., and S. Foner,
Phys. Rev. Lett. {\bf 77}, 394 (1996). \url{https://doi.org/10.1103/PhysRevLett.77.394}

\bibitem {Sousa} M. H. Sousa, E. Hasmonay, J. Depeyrot, F.A . Tourinho, J.-C. Bacri, E. Dubois, R. Perzynski, and Yu. L. Raikher, J. Magn. Magn. Mater. {\bf 242-245}, 572-574 (2002). \url{https://doi.org/10.1016/S0304-8853(01)01122-2}

\bibitem{Pastukh} O. Pastukh, D. Kuz\'ma, and P. Zielin\'ski, Crystals {\bf 13}, 1617 (2023). \url{https://doi.org/10.3390/cryst13121617}

\bibitem{Egbu} J. Egbu, P. R. Ohodnicki, J. P. Baltrus, A. Talaat, R. F. Wright, and M. E. McHenry, J. Alloys Compd. {\bf 912}, 165155 (2022). \url{https://doi.org/10.1016/j.jallcom.2022.165155}

\bibitem{Heo} S. Heo, S. H. Cho, C. J. Dahlman, A. Agrawal, and D. J. Milliron, ACS Energy Lett. {\bf 5}, 2662 (2020). \url{https://doi.org/10.1021/acsenergylett.0c01236}

\bibitem{Gomonay} O. Gomonay, S. Kondovych and V. Loktev, J. Magn. Magn. Mater {\bf 354}, 125-135 (2014). \url{https://doi.org/10.1016/j.jmmm.2013.11.003}

\bibitem{Zou} W. Zou, J. Dong, M. Ji, B. Wang, Y. Li, S. Yin, H. Li, J. Xia, ACS Appl. Nano Mater. {\bf 6}, 4309-4318 (2023). \url{https://doi.org/10.1021/acsanm.2c05448}

\bibitem{He} X. He, Y. He, C. Wang, B. Zhu, A. Liu, and Hui Tang, J. Phys. Chem. Solids {\bf 171}, 111028 (2022). \url{https://doi.org/10.1016/j.jpcs.2022.111028}

\bibitem{Danilyuk} A. L. Danilyuk, A. V. Kukharev, and S. L. Prischepa, IEEE Trans. Magn. {\bf 58}, 1-5 (2022), \url{https://doi.org/10.1109/TMAG.2021.3102403}

\bibitem{salazar} G. Salazar-Alvarez, J. Qin, V. Šepel\'{a}k, I. Bergmann, M. Vasilakaki, K. N. Trohidou, J. D. Ardisson, W. A. A. Macedo, M. Mikhaylova, M. Muhammed, M. D. Bar\'{o} and J. Nogu\'{e}s, J. Am. Chem. Soc. {\bf 130}, 13234-13239 (2008). \url{https://doi.org/10.1021/ja0768744}

\bibitem{Pierce} D. T. Pierce, J. Unguris, R.J. Celotta, M.D. Stiles, J. Magn. Magn. Mater. {\bf 200}, 290-321 (1999). \url{https://doi.org/10.1016/S0304-8853(99)00319-4}

\bibitem{Razouk} A. Razouk, M. Sahlaoui, S. Eddahri, J. Supercond. Nov. Magn. {\bf 30}, 425–430 (2017). \url{https://doi.org/10.1007/s10948-016-3735-4}

\bibitem{wanga} G. Wanga, Y. Xub, P. Qiana, Y. Sua, Comput. Mater. Sci. {\bf 173} (2020) 109. \url{https://doi.org/10.1016/j.commatsci.2019.109416}

\bibitem{goyal} M. Goyal, Vishal Goyal, {\bf 95}, 99 (2021). \url{https://doi.org/10.1007/s12043-021-02127-8}

\bibitem{agarwal} M. Agarwal and E. G. Mishchenko, Phys. Rev. B {\bf 95}, 075411 (2017). \url{https://doi.org/10.1103/PhysRevB.95.075411}

DOI: https://doi.org/10.1103/PhysRevB.95.075411

\bibitem{kondo} J. Kondo, Prog. Theor. Phys. {\bf 32}, 37–49 (1964). \url{ https://doi.org/10.1143/PTP.32.37}

\bibitem{pandey} S. Pandey, S. Das, B. Kamble, S., D. Singh, R. Ray, and A. Singh, Phys. Rev. B {\bf 22}, 134447 (2008). \url{https://doi.org/10.1103/PhysRevB.77.134447}

\bibitem{dks1} D. K. Singh, B. Kamble, and A. Singh, J. Phys.: Condens. Matter {\bf 22}, 396001 (2010). \url{https://doi.org/10.1088/0953-8984/22/39/396001}

\bibitem{dks2} D. K. Singh and A. Singh, Phys. Rev. B {\bf 88}, 144410 (2013). \url{https://doi.org/10.1103/PhysRevB.88.144410}

\bibitem{anderson} P. W. Anderson, Phys. Rev. {\bf 79}, 350 (1950). \url{https://doi.org/10.1103/PhysRev.79.350}

\bibitem{harun1} H. A. Rashid, D. K. Singh, Phys. Rev. B {\bf 107}, 125139 (2023). \url{https://doi.org/10.1103/PhysRevB.107.125139}

\bibitem{harun2} H. A. Rashid and D. K. Singh, SciPost Phys. {\bf 16}, 107 (2024). \url{https://doi.org/10.21468/SciPostPhys.16.4.107}

\bibitem{liu} C. Liu, X. Lu, P. Dai, R. Yu, and Q. Si, Phys. Rev. B {\bf 101}, 024510 (2020). \url{https://doi.org/10.1103/PhysRevB.101.024510}

\bibitem{Zhang} Y. Zhang, B. Wang, Y. Guo, Q. Li, and J. Wang, Comp. Mater. Sci. {\bf 197}, 110638 (2021). \url{https://doi.org/10.1016/j.commatsci.2021.110638}

\bibitem{patel} N. D. Patel, A. Mukherjee, N. Kaushal, A. Moreo, E. Dagotto, Phys. Rev. Lett. {\bf 119}, 086601 (2017). \url{https://doi.org/10.1103/PhysRevLett.119.086601}

\bibitem{carvalho} D. C. Carvalho, A. S. T. Pires, L. A. S. Mól, J. Magn. Magn. Mater {\bf 407} 341 (2016). \url{https://doi.org/10.1016/j.jmmm.2016.02.001}

\bibitem{meyer} R. Meyer, G. dos Santos, R. Aparicio, E. M. Bringa, H. M. Urbassek, Comput. Cond. Matt. {\bf 31} e00662 (2022). \url{https://doi.org/10.1016/j.cocom.2022.e00662}

\end{thebibliography}
\end{document}